\documentclass[11pt,english,a4paper]{llncs}
\usepackage{fullpage}

\usepackage{amsmath}
\usepackage{amssymb}
\usepackage{color}
\usepackage{subfigure}
\usepackage{graphicx}
\usepackage{grffile}

\begin{document}
\mainmatter              

\title{Self-Indexing Based on LZ77}

\titlerunning{Self-Index Based on LZ77}  

\author{Sebastian Kreft \and Gonzalo Navarro}
\authorrunning{S. Kreft and G. Navarro} 
\institute{Dept. of Computer Science, University of Chile, Santiago, Chile.
\email{\{skreft,gnavarro\}@dcc.uchile.cl}
}

\maketitle              

\begin{abstract}
\vspace*{-5mm}
We introduce the first self-index based on the Lempel-Ziv 1977 compression
format (LZ77). It is particularly competitive for highly repetitive text 
collections such as sequence databases of genomes of related species, software 
repositories, versioned document collections, and temporal text databases.
Such collections are extremely compressible but classical self-indexes fail
to capture that source of compressibility. Our self-index takes in practice
a few times the space of the text compressed with LZ77 (as little as 2.6 
times), extracts 1--2 million characters of the text per second, and finds 
patterns at a rate of 10--50 microseconds per occurrence. It is smaller (up 
to one half) than the best current self-index for repetitive collections, and 
faster in many cases.
\end{abstract}

\section{Introduction and Related Work}

Self-indexes \cite{NM07} are data structures that represent a text collection
in compressed form, in such a way that not only random access to the text is
supported, but also indexed pattern matching. Invented in the past decade,
they have been enormously successful to drastically reduce the space burden
posed by general text indexes such as suffix trees or arrays. Their compression
effectiveness is usually analyzed under the $k$-th order entropy model
\cite{Man01}: $H_k(T)$ is the $k$-th order entropy of text $T$,
a lower bound to the bits-per-symbol compression achievable by any statistical
compressor that models symbol probabilities as a function of the $k$ symbols
preceding it in the text. There exist self-indexes able to represent a text 
$T_{1,n}$ over alphabet $[1,\sigma]$, within $nH_k(T)+o(n\log\sigma)$ bits of
space for any $k \le \alpha\log_\sigma n$ and constant $\alpha<1$
\cite{GGV03,FMMN07}.

This $k$-th order entropy model is adequate for many practical text collections.
However, it is not a realistic lower bound model for a kind of collections
that we call {\em highly repetitive}. This is formed by sets of strings that 
are mostly near-copies of each other. For example, versioned document
collections store all the history of modifications of the documents. Most new
versions are minor edits of a previous version. Good examples are the Wikipedia
database and the Internet archive. Another example are software repositories,
which store all the versioning history of software pieces. Again, except for
major releases, most versions are minor edits of previous ones. In this case 
the versioning has a tree structure more than a linear sequence of versions. 
Yet another example comes from bioinformatics. Given the sharply decreasing 
sequencing costs, large sequence databases of individuals of the same or 
closely related species are appearing. The genomes of two humans, for example, 
share 99.9\% to 99.99\% of their sequence. 
No clear structure such as a versioning tree is apparent in the general case.

If one concatenates two identical texts, the statistical structure of the
concatenation is almost the same as that of the pieces, and thus the $k$-th
order entropy does not change. As a consequence, some indexes that are exactly
tailored to the $k$-th order entropy model \cite{GGV03,FMMN07} are insensitive 
to the repetitiveness of the text. M\"akinen et al.\ \cite{MNSV08,MNSV09} 
found that even the self-indexes that can compress beyond the $k$-th order 
entropy model \cite{Sad03,Nav04} failed to capture most of the repetitiveness 
of such text collections.

Note that we are not aiming simply at {\em representing} the text collections
to offer {\em extraction} of individual documents. This is relatively simple
as it is a matter of encoding the edits with respect to some close sampled
version; more sophisticated techniques have been however proposed for this
goal \cite{KBSCZ09,KPZ10,KN10}. Our aim is more ambitious: self-indexing the
collection means providing not only access but indexed searching, just as if
the text was available in plain form. Other restricted goals such as compressing
the inverted index (but not the text) on natural-language text collections
\cite{HZS10} 
or indexing text $q$-grams and thus fixing the pattern 
length in advance \cite{CFMPN10} have been pursued as well.

M\"akinen et al.\ \cite{MNSV08,MNSV09} demonstrated that repetitiveness in the
text collections translates into {\em runs} of equal letters in its
Burrows-Wheeler transform \cite{BW94} or runs of successive values in the
$\Psi$ function \cite{GV00}. Based on this they engineered variants of 
FM-indexes \cite{FMMN07} and Compressed Suffix Arrays (CSAs) \cite{Sad03} that 
take advantage of repetitiveness. Their best structure, the Run-Length CSA 
(RLCSA) still stands as the best general-purpose self-index, despite of some 
preliminary attempts of self-indexing based on grammar compression 
\cite{CFMPN10}.

However, M\"akinen et al.\ showed that their new self-indexes
were very far (by a factor of 10) from the space that can be achieved by a
compressor based on the Lempel-Ziv 1977 format (LZ77) \cite{ZL77}. They showed
the runs model is intrinsically inferior to the LZ77 model to capture 
repetitions. The LZ77 compressor is particularly able to capture 
repetitiveness, as it parses the text into consecutive maximal {\em phrases} 
so that each phrase appears earlier in the text. A self-index based on LZ77 
was advocated as a very promising alternative approach to the problem.

Designing a self-index based on LZ77 is challenging. Even accessing 
LZ77-compressed text at random is a difficult problem, which we partially
solved \cite{KN10} with the design of a variant called LZ-End, which compresses
only slightly less and gives some time guarantees for the access time. There
exists an early theoretical proposal for LZ77-based indexing by K\"arkk\"ainen
and Ukkonen \cite{KU96,Kar99}, 
but it requires to have the text in plain form and has never been implemented.
Although it guarantees an index whose size is of the same order of the LZ77
compressed text, the constant factors are too large to be practical. 
Nevertheless, that index was the first general compressed index in the 
literature and is the predecessor of all the Lempel-Ziv indexes that 
followed \cite{Nav04,FM05,RO08}. These indexes have used variants of the LZ78 
compression format \cite{ZL78}, which is more tractable but still too weak to
capture high repetitiveness \cite{MNSV08}.

In this paper we face the challenge of designing the first self-index based on
LZ77 compression. Our self-index can be seen as a modern variant of 
K\"arkk\"ainen and Ukkonen's LZ77 index, which solves the problem of not having
the text at hand and also makes use of recent compressed data structures.
This is not trivial at all, and involves designing new solutions to some
subproblems where the original solution \cite{KU96} was too space-consuming.
Some of the solutions might have independent interest.

Our resulting index is competitive in theory and in
practice. Let $n'$ be the number of phrases of the LZ-parsing of $T_{1,n}$.
Then a Lempel-Ziv compressor output has size $|LZ| = n'(2\log n + \log\sigma)$
(our logarithms are base 2 by default). The size of our LZ77 self-index is 
at most $2|LZ|+ o(|LZ|)$. It can determine the existence of pattern $p_{1,m}$
in $T$ in time $O(m^2h+m\log n')$, where $h \le n'$ is a measure of the nesting
of the parsing (i.e., how many times a character is transitively copied),
usually a small number. After this check, each occurrence is reported in
time $O(\log n' + \delta\log\delta)$, where $\delta$ is another usually small 
number depending on the nesting of the parsing.

We implemented our self-index over LZ77 and LZ-End parsings, and compared it 
with the state of the art on a number of real-life repetitive collections 
consisting of Wikipedia versions, versions of public software, periodic 
publications, and DNA sequence collections. We have left a public repository 
with those repetitive collections in 
{\tt http://pizzachili.dcc.uchile.cl/ repcorpus.html}, so that 
standardized comparisons are possible. Our implementations and those of the 
RLCSA are also available in there.

Our experiments show that in practice the smallest-space variant of our index 
takes 2.5--4.0 times the space of a LZ77-based compressor, it can extract 1--2 
million characters per second, and locate each occurrence of a pattern of 
length 10 in 10--50 microseconds. Compared to the state of the art (RLCSA), our 
self-index always takes less space, less than a half on our DNA and 
Wikipedia corpus. Searching for short patterns is faster than on the 
RLCSA. On longer patterns our index offers competitive space/time trade-offs.

\section{Direct Access to LZ-Compressed Texts}

Let us first recall the classical LZ77 parsing \cite{ZL77}, as well as the 
recent LZ-End parsing \cite{KN10}. This involves defining what is a phrase and
its source, and the number $n'$ of phrases.

\begin{definition}[\cite{ZL77}]
The \emph{LZ77 parsing} of text $T_{1,n}$ is a sequence $Z[1,n']$ of
\emph{phrases} such that $T = Z[1] Z[2] \ldots Z[n']$, built as follows. 
Assume we
have already processed $T_{1,i-1}$ producing the sequence $Z[1,p-1]$. Then, we
find the longest prefix $T_{i,i'-1}$ of $T_{i,n}$ which occurs in
$T_{1,i-1}$, set $Z[p] = T_{i,i'}$ and continue with $i = i'+1$. The
occurrence in $T_{1,i-1}$ of prefix $T_{i,i'-1}$ is called the \emph{source} of 
the phrase $Z[p]$.
\end{definition}

\begin{definition}[\cite{KN10}]
The \emph{LZ-End parsing} of text $T_{1,n}$ is a sequence $Z[1,n']$ of
\emph{phrases} such that $T = Z[1] Z[2] \ldots Z[n']$, built as follows. 
Assume we
have already processed $T_{1,i-1}$ producing the sequence $Z[1,p-1]$. Then, we
find the longest prefix $T_{i,i'-1}$ of $T_{i,n}$ that is a suffix of 
$Z[1] \ldots Z[q]$ for some $q<p$,
set $Z[p] = T_{i,i'}$ and continue with $i = i'+1$. 
\end{definition}

\begin{figure}[tb]

\subfigure[The LZ77 parsing of the string \texttt{`alabar\_a\_la\_alabarda\$'},
showing the sources of each phrase on top. On the bottom, bitmap $B$ marks
the ends of phrases, the bitmap $S$ marks the starting positions of sources,
and the permutation $P$ connects phrases to sources.]{
\begin{minipage}{0.45\textwidth}
\vspace*{-7cm}
\includegraphics[width=\textwidth]{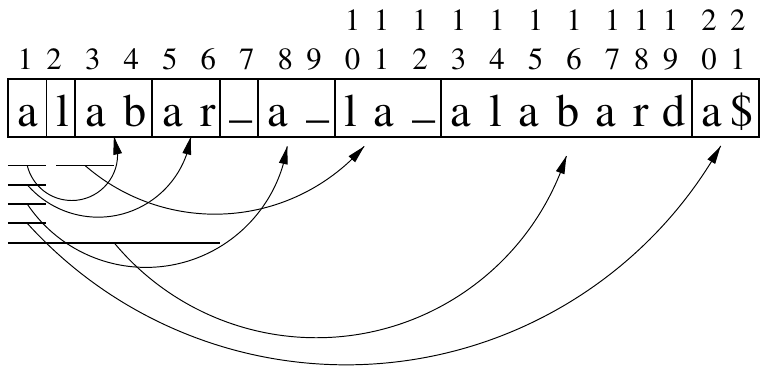}

\includegraphics[width=\textwidth]{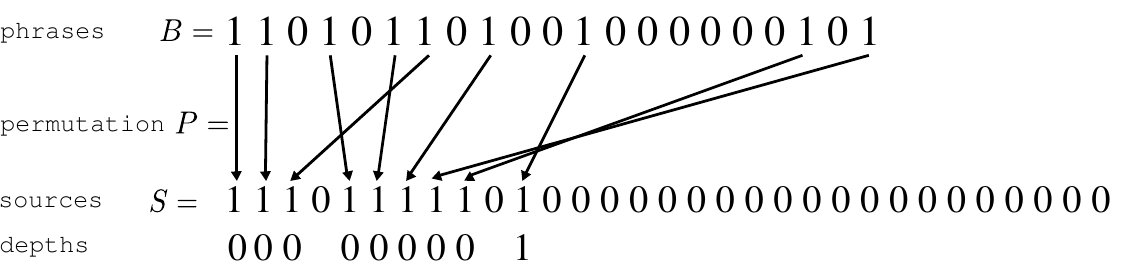}
\end{minipage} 
\label{fig:parsing}
}
\hfill
\subfigure[Top: The sparse suffix trie. The black node is the one we arrive at
when searching for \texttt{`la'}, and the gray leaves of its subtree represent 
the phrases that start with \texttt{`la'}. Left: The reverse trie for 
the string. The gray leaf is the node at which we stop searching for 
\texttt{`a'}. Bottom: The range structure for the string. The gray circle 
marks the only primary occurrence of the pattern \texttt{`ala'} (it is the only dot in the range defined by the gray nodes).]{
\includegraphics[width=0.50\textwidth]{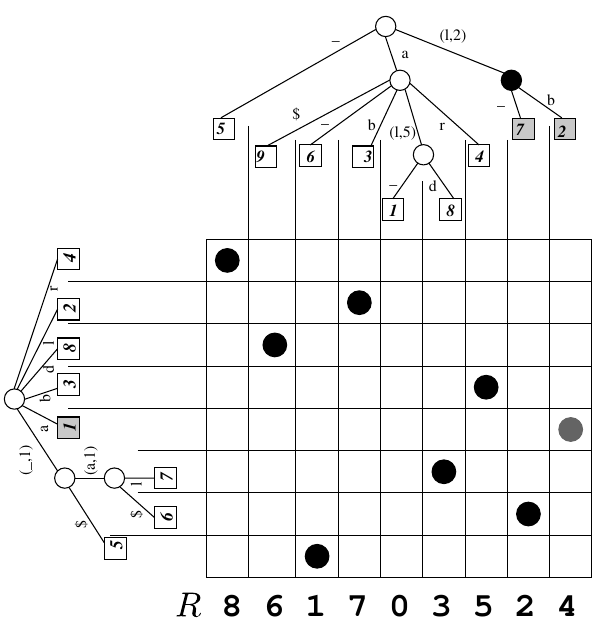}
\label{fig:primary}
}
\caption{Our self-index structure over the example text
$T=\texttt{`alabar\_a\_la\_alabarda\$'}$ and part of the process of searching 
for $p=\texttt{`ala'}$.}
\end{figure}

We will store $Z$ in a particular way that enables efficient extraction of
any text substring $T_{s,e}$. This is more complicated than in our previous
proposal \cite{KN10} because these structures will be integrated into the
self-index later. First, the last characters of the phrases, $T_{i'}$ of
$Z[p] = T_{i,i'}$, are stored in a string $L_{1,n'}$. Second, we 
set up a bitmap $B_{1,n}$ that will mark with a 1 the ending positions of
the phrases in $T_{1,n}$ (or, alternatively, the positions where the
successive symbols of $L$ lie in $T$). Third, we store a bitmap $S_{1,n+n'}$
that describes the structure of the sources in $T$, as follows. We traverse
$T$ left to right, from $T_1$ to $T_n$. At step $i$, if there are $k$ sources
starting at position $T_i$, we append $1^k0$ to $S$ ($k$ may be zero). 
Empty sources (i.e., $i=i'$ in $Z[p]=T_{i,i'}$) are assumed
to lie just before $T_1$ and appended at the beginning of $S$, followed by a 0.
So the 0s in $S$ correspond to text positions, and the 1s correspond to 
the successive sources, where we assume that those that start at the same 
point are sorted by shortest length first. Finally, we store a permutation 
$P[1,n']$ that maps targets to sources, that is, $P[i]=j$ means that the source 
of the $i$th phrase starts at the position corresponding to the $j$th 1 in $S$.
Fig.~\ref{fig:parsing} gives an example.

The bitmaps $B_{1,n}$ and $S_{1,n+n'}$ are sparse, as they have only $n'$ bits
set. They are stored using a compressed representation \cite{RRR02} so that
each takes $n'\log\frac{n}{n'} + O(n') + o(n)$ bits, and rank/select queries
are answered in constant time: $rank_b(B,i)$ is the number of occurrences of
bit $b$ in $B_{1,i}$, and $select_b(B,j)$ is the position in $B$ of the $j$th
occurrence of bit $b$ (similarly for $S$). The $o(n)$ term, the only one that
does not depend linearly on $n'$, can disappear at the cost of increasing the
time for $rank$ to $O(\log\frac{n}{n'})$ \cite{OS07}. Finally, permutations
are stored using a representation \cite{MRRR03} that computes $P[i]$ in
constant time and $P^{-1}[j]$ in time $O(l)$, using $(1+1/l)n'\log n'+O(n')$ 
bits of space. We use parameter $l=\log n'$. Thus our total space is
$n'\log n + n'\log \frac{n}{n'} + n'\log\sigma + O(n') + o(n)$ bits.

To extract $T_{s,e}$ we proceed as follows. We compute
$s' = rank_1(B,s-1)+1$ and $e'=rank_1(B,e)$
to determine that we must extract characters from phrases $s'$ to $e'$.
For all phrases except possibly $e'$ (where $T_{s,e}$ could end before its
last position) we have their last characters in $L[s',e']$. For all the
other symbols, we must go to the source of each phrase of length more than one
and recursively extract its text: to extract the rest of phrase $s' \le k \le
e'$, we compute its length as $l=select_1(B,k)-select_1(B,k-1)$ (except for
$k=e'$, where the length is $l=e-select_1(B,k-1)$) and its starting position 
as $t=rank_0(S,select_1(S,P[k])) = select_1(S,P[k])-P[k]$. Thus to obtain the 
rest of the characters of phrase $k$ we recursively extract $T_{t,t+l-1}$

On LZ-End this method takes time $O(e-s+1)$ if $e$ coincides with the end of
a phrase \cite{KN10}. In general, a worst-case analysis \cite{KN10} yields
extraction time $O(e-s+h)$ for LZ-End and $O(h(e-s+1))$ for LZ77, where $h$
is a measure of how nested is the parsing. 

\begin{definition}
Let $T = Z[1] Z[2] \ldots Z[n']$ be a LZ-parsing of $T_{1,n}$. Then the {\em
height} of the parsing is defined as $h = \max_{1\le i\le n}C[i]$, where $C$
is defined as follows. Let $Z[i] = T_{a,b}$ be a phrase whose source is 
$T_{c,d}$. Then $C[b]=1$ and $C[k] = C[(k-a)+c]+1$ for $a \le k < b$.
\end{definition}

That is, $h$ measures how many times a character is transitively copied in $Z$.
While in the worst case $h$ can be as large as $n'$, it is usually a small 
value. It is limited by the longest length of a phrase \cite{Kre10}, thus on
a text coming from a Markovian source it is $O(\log_\sigma n)$. On our 
repetitive collection corpus $h$ is between 22 and 259 for LZ-End, and between
22 and 1003 for LZ77. Its average values, on the other hand, are 5--25 on LZ-End
and 5--176 on LZ77.

\paragraph{Implementation considerations.} As bitmaps $B$ and $S$ are very
sparse in highly repetitive collections, we opted for $\delta$-encoding the
distances between the consecutive 1s, and adding a sampling where we store the
absolute values and position in the $\delta$-codes of every $s$th bit, where
$s$ is the sampling rate. So $select$ consists in going to the previous sample
and decoding at most $s$ $\delta$-codes, whereas $rank$ requires a previous
binary search over the samples.

\section{Pattern Searches}

Assume we have a text $T$ of length $n$, which is partitioned into $n'$ phrases 
using a LZ77-like compressor. Let $p_{1,m}$ be a search pattern. We call 
\emph{primary occurrences} of $p$ those covering more than one phrase or ending
at a phrase boundary; and \emph{secondary occurrences} the others. For example,
in Fig.~\ref{fig:parsing}, the occurrence of \texttt{`lab'} starting at 
position 2 is primary as it spans two phrases. The second occurrence, starting 
at position 14, is secondary. 

We will find first the primary occurrences, and those will be used to 
recursively find the secondary ones (which, in turn, will be used to find 
further secondary occurrences).

\subsection{Primary Occurrences}

Each primary occurrence can be split as $p=p_{1,i}\,p_{i+1,m}$, where the left 
side $p_{1,i}$ is a nonempty suffix of a phrase and the (possibly empty) right 
side $p_{i+1,m}$ is 
the concatenation of zero or more consecutive phrases plus a prefix of the 
next phrase. To find primary occurrences we partition the pattern into two in 
every possible way. Then, we search for the left part in the suffixes of the 
phrases and for the right part in the prefixes of the suffixes of $T$ starting 
at phrase boundaries. Then, we find which pairs of left and right occurrences 
are concatenated, thus representing actual primary occurrences of $p$.

\paragraph{Finding the Right Part of the Pattern.}

To find the right side $p_{i+1,m}$ of the pattern we use a suffix trie that 
indexes all the suffixes of $T$ starting at the beginning of a phrase. In the 
leaves of the trie we store the identifiers of the phrases where the
corresponding suffixes start. Conceptually, the identifiers form an array $id$ 
that stores the phrase identifiers in lexicographic order of their suffixes. 
As we see later, we do not need to store $id$ explicitly.

We represent the suffix trie as a Patricia tree \cite{Mor68}, encoded using
a succinct representation for labeled trees called {\em dfuds} \cite{BDMRRR05}.
As the trie has at most $2n'$ nodes, the succinct representation requires at 
most $2n'\log\sigma + O(n')$ bits. It supports a large number of operations in
constant time, such as going to a child labeled $c$, going to the leftmost and
rightmost descendant leaf, etc.
To search for $p_{i+1,m}$ we descend through the tree using the next character
of the pattern, skip as many characters as the skip of the branch indicates,
and repeat the process until determining that $p_{i+1,m}$ is not in the set or
reaching a node or an edge, whose leftmost and rightmost subtree leaves define 
the interval in array $id$ whose suffixes start with $p_{i+1,m}$.
Fig.~\ref{fig:primary} shows on top this trie, shading the range [8,9] of
leaves found when searching for $p_{i+1,m} = \texttt{`la'}$.

Recall that, in a Patricia tree, after searching for the positions we need to 
check if they are actually a match, as some characters are not checked because 
of the skips. 

Instead of doing the
check at this point, we defer it for later, when we connect both searches.

We do not explicitly store the skips, as they can be computed from the trie
and the text. Given a node in the trie corresponding to a string of length
$l$, we go to the leftmost and rightmost leaves and extract the corresponding 
suffixes since their $(l+1)$th symbols. The number $s$ of symbols they share 
since that position is the skip. This takes $O(sh)$ time both for LZ77 and 
LZ-End, since the extraction is from left to right and we have to extract one 
character at a time until they differ. Thus, the total time for extracting the 
skips as we descend is $O(mh)$.

\paragraph{Finding the Left Part of the Pattern.}

We have another Patricia trie that indexes all the reversed phrases,
stored in the same way as the suffix trie. 
To find the left part of the pattern in the text we search for 
$(p_{1,i})^{rev}$ in this trie. The array that stores the leaves of the trie 
is called $rev\_id$ and is stored explicitly. The total space is at most
$n'\log n' + 2n'\log\sigma + O(n')$ bits.
Fig.~\ref{fig:primary} shows this trie on the left, with the result of 
searching for a left part $p_{1,i} = \texttt{`a'}$.

\paragraph{Connecting Both Searches.}

Actual occurrences of $p$ are those formed by a phrase $rev\_id[j]=k-1$ and
the following one $id[i]=k$, so that $j$ and $i$ belong to the lexicographical
intervals found with the tries.
To find those we use a $n' \times n'$ range structure that connects the 
consecutive phrases in both trees. If $id[i]=k$ and $rev\_id[j]=k-1$, 
the structure holds a point in $(i,j)$. 

The range structure is represented 
compactly using a wavelet tree \cite{GGV03,MN07}, which requires 
$n'\log n' + O(n'\log\log n')$ bits. This can be reduced to $n'\log n'+O(n')$
\cite{Pat08}. The wavelet tree stores the sequence 
$R[1,n']$ so that $R[i]=j$ if $(i,j)$ is a point (note there is only one $j$
per $i$ value). In $O(\log n')$ time it can compute $R[i]$, as
well as find all the $occ$ points in a given orthogonal range in time
$O((occ+1)\log n')$. With such an orthogonal range search for the intervals of 
leaves found in both trie searches, the wavelet tree gives us all the primary 
occurrences. It also computes any $id[i] = rev\_id[R[i]]+1$ in
$O(\log n')$ time, thus we do not need to store $id$.

Fig.~\ref{fig:primary} gives an example, showing sequence $R$ at the bottom.
It also shows how we find the only primary occurrence of $p=\texttt{`ala'}$
by partitioning it into \texttt{`a'} and \texttt{`la'}.

At this stage we also verify that the answers returned by the searches in 
the Patricia trees are valid. It is sufficient to extract the text of one
of the occurrences reported and compare it to $p$, to determine either that 
all or none of the answers are valid, by the Patricia tree properties.

Note that the structures presented up to now are sufficient to determine whether
the pattern exists in the text or not, since $p$ cannot appear if it does not 
have primary occurrences. If we have to report the $occ$ occurrences, instead,
we use bitmap $B$: An occurrence with partition $p_{1,i}$ and $p_{i+1,m}$
found at $rev\_id[j]=k$ is to be reported at text position $select_1(B,k)-i+1$.

Overall, the data structures introduced in this section add up to
$2n'\log n' + 4n'\log\sigma + O(n')$ bits. The $occ$ primary 
occurrences are found in time $O(m^2h + m\log n' + occ\log n')$.

\paragraph{Implementation Considerations.}

As the average value for the skips is usually very low and computing them from 
the text phrases is slow in practice, we actually store the skips using
\emph{Directly Addressable Codes} \cite{BLN09}. These
allow storing variable-length codes while retaining fast direct access. 
In this case arrays $id$ and $rev\_id$ are only accessed for reporting
the occurrences.

We use a practical {\em dfuds} implementation \cite{ACNS10} that binary
searches for the child labeled $c$, as the theoretical one \cite{BDMRRR05}
uses perfect hashing.

Instead of storing the tries we can do a binary search over the $id$ or
$rev\_id$ arrays. This alternative modifies the 
complexity of searching for a prefix/suffix of $p$ to $O(mh\log n')$ for LZ77 
or $O((m+h)\log n')$ for LZ-End.

Independently, we could store explicitly array $id$, instead of accessing it
through the wavelet tree. Although this alternative increases the space usage 
of the index and does not improve the complexity, it gives an interesting 
trade-off in practice.

\subsection{Secondary Occurrences}
\label{sec:secondary}

Secondary occurrences are found from the primary occurrences and, recursively, from other previously discovered secondary occurrences. 
The idea is to locate all sources covering the occurrence and then finding
their corresponding phrases. Each discovered copy is reported and recursively
analyzed for sources containing it.

For each occurrence found $T_{i,i+m-1}$, we find the position \emph{pos} of 
the 0 corresponding to its starting position in bitmap $S$,
$pos = select_0(S,i)$. Then we consider all the 1s to the left of \emph{pos},
looking for sources that start before the occurrence. For each such $S[j]=1$,
$j \le pos$, the source starts in $T$ at $t=rank_0(S,j)$ and is the $s$th source,
for $s = rank_1(S,j)$. Its corresponding phrase is $f = P^{-1}[s]$, which 
starts at text position $c=select(B,f-1)+1$. Now we can compute the length of 
the source, which is the length of its phrase minus one, 
$l=select_1(B,f)-select_1(B,f-1)-1$. Finally, if $T_{t,t+l-1}$ covers the
occurrence $T_{i,i+m-1}$, then this occurrence has been copied to
$T_{c+i-t,c+i-t+m-1}$, where we report a secondary occurrence and recursively
find sources covering it. The time per occurrence reported is
dominated by that of computing $P^{-1}$, $O(\log n')$. 

Consider the only primary occurrence of pattern \texttt{`la'} 
starting at position 2 in our example text. We find the third 0 in the bitmap 
of sources at position 12. Then we consider all ones starting from position 11 
to the left. The 1 at position 11 maps to a phrase of length 2 that 
covers the occurrence, hence we report an occurrence at position 10. The second
1 maps to a phrase of length 6 that also covers the occurrence, thus we report 
another occurrence at position 15. The third 1 maps to a phrase of length 1, 
hence it does not cover the occurrence and we do not report it. We proceed 
recursively for the occurrences found at positions 10 and 15.

Unfortunately we do not know when to stop looking for 1s to the left in $S$. 
Stopping as soon as we find the first source not covering the occurrence works 
only when no source contains another. K\"arkk\"ainen \cite{Kar99} proposes a 
couple of solutions to this problem, but none is satisfactory in practice.

However, one involves a concept of {\em levels}, which we use here in a 
different way.

\begin{definition}
\label{def:depth_source}
Source $s_1=[l_1,r_1]$ is said to {\em cover} source $s_2=[l_2,r_2]$ if 
$l_1 < l_2$ and $r_1\ge r_2$. 

Let $cover(s)$ be the set of sources covering a source $s$. Then the 
\emph{depth} of source $s$ is defined as $depth(s)=0$ if $cover(s)=\emptyset$, 
and $depth(s)=1+\max_{s' \in cover(s)}depth(s')$ otherwise. We define 
$depth(\varepsilon)=0$. Finally, we call $\delta$ the maximum depth in the 
parsing.
\end{definition}

In our example, the four sources \texttt{`a'} and the source \texttt{`alabar'} 
have depth zero, as all of them start at the same position. Source 
\texttt{`la'} has depth 1, as it is contained by source \texttt{`alabar'}.

We slightly modify the process for traversing $S$ to the left of $pos$.
When we find a source not covering the occurrence, we look for 
its depth $d$ and then consider to the left only sources with depth $d'<d$, as 
those at depth $\ge d$ are guaranteed not to contain the occurrence. This 
works because sources to the left with the same depth $d$ will end before the 
current source, and deeper sources to the left will be contained in those of 
depth $d$. Thus for our traversal we need to solve a subproblem we call 
$prevLess(D,s,d)$: Let $D[1,n']$ be the array of depths of the sources; 
given a position $s$ and a depth $d$, we need to find the largest $s'<s$ such 
that $D[s'] < d$. 

\paragraph{Prev-Less Data Structure.}

We represent $D$ using a wavelet tree \cite{GGV03}. This
time we need to explain its internal structure. The wavelet tree is a balanced
tree where each node represents a range of the alphabet $[0,\delta]$. The root
represents the whole range and each leaf an individual alphabet member. Each
internal node has two children that split its alphabet range by half. Hence the
tree has height $\lceil \log (1+\delta) \rceil$. At the root node, the tree
stores a bitmap aligned to $D$, where a 0 at position $i$ means that $D[i]$ is
a symbol belonging to the range of the left child, and 1 that it belongs to
the right child. Recursively, each internal node stores a bitmap that refers to
the subsequence of $D$ formed by the symbols in its range. All the bitmaps are
preprocessed for rank/select queries, needed for navigating the tree. 
The total space is $n'\log\delta + O(n')$.

We solve $prevLess(D,s,d)$ as follows. We descend on the wavelet tree towards
the leaf that represents $d-1$. If $d-1$ is to the left of the current node, then
no interesting values can be stored in the right child. So we recursively 
continue in the left subtree, at position $s'=rank_0(V,s)$, where $V$ is the 
bitmap of the current node. Otherwise we descend to the right child, and the 
new position is $s'=rank_1(V,s)$. In this case, however, the answer could be 
at the left child. Any value stored at the left child is $<d$, so we are 
interested in the rightmost before position $s$. 
Hence $v_0 = select_0(V,rank_0(V,s-1))$ is the last relevant position with
a value from the left subtree. We find, recursively, the best answer $v_1$ from
the right subtree, and return $\max(v_0,v_1)$. When the recursion ends at a
leaf we return with answer $-1$.
The running time is $O(\log \delta)$. 

Using this operation we proceed as follows. We keep track of the smallest depth
$d$ that cannot cover an occurrence; initially $d=\delta+1$. We start considering 
source $s$. Whenever $s$ does not cover the occurrence, we set $d=D[s]$ and 
move to $s'=prevLess(D,s,d)$. When $s$ covers the occurrence, we
report it and move to $s'=prevLess(D,s,d)$.

In the worst case the first source is at depth $\delta$ and then we traverse
level by level, finding in each previous source that however does not contain
the occurrence. Therefore the overall time is $O(occ(\log n'+\delta\log\delta))$
to find $occ$ secondary occurrences.

\paragraph{Final Bounds.}
Overall our data structure requires $2n'\log n + n'\log n' + n'\log\delta + 
5n'\log\sigma + O(n') + o(n)$ bits, where the last term can be removed as 
explained (multiplying times by $O(\log n)$).
As the Lempel-Ziv compressor output has 
$n'(2\log n + \log\sigma)$ bits, the index is asymptotically at most twice the
size of the compressed text (for $\log\sigma=o(\log n)$; 3 times otherwise).
In practice $\delta$ is much smaller: it is also
limited by the maximum phrase length, thus on Markovian sources it is 
$O(\log_\sigma n)$, and in our test collections it is at most 46 and on 
average 2--4.

Our time to locate the $occ$ occurrences of $p_{1,m}$ is $O(m^2h+m\log n' +
occ(\log n' + \delta\log\delta))$.

%
%
\section{Experimental Evaluation}

From the testbed in {\tt http://pizzachili.dcc.uchile.cl/repcorpus.html} we
have chosen four real collections representative of distinct applications:
{\tt Cere} (37 DNA sequences of Saccharomyces Cerevisiae),
{\tt Einstein} (the version of the Wikipedia article on Albert Eintein up to
Jan 12, 2010), {\tt Kernel} (the 36 versions 1.0.x and 1.1.x of the Linux
Kernel), and {\tt Leaders} (pdf files of the CIA World Leaders report,
from Jan 2003 to Dec 2009, converted with {\tt pdftotext}).

We have studied 5 variants of our indexes, from most to least space consuming:
(1) with suffix and reverse trie; (2) binary search on explicit $id$ array 
and reverse trie; (3) suffix trie and binary search on $rev\_id$; (4) binary 
search on explicit $id$ array and on $rev\_id$; (5) binary search on 
implicit $id$ and on $rev\_id$. In addition we test parsings LZ77 and LZ-End,
so for example LZ-End$_3$ means variant (3) on parsing LZ-End.

Table~\ref{tab:space} gives statistics about the texts, with the compression
ratios achieved with a good Lempel-Ziv compressor ({\tt p7zip}, 
\verb'www.7-zip.org'), grammar compressor ({\tt repair}, 
\verb'www.cbrc.jp/' \verb'~rwan/en/restore.html'), Burrows-Wheeler compressor
({\tt bzip2}, \verb'www.bzip.org'), and statistical high-order compressor
({\tt ppmdi}, \verb'pizzachili.dcc.uchile.cl/utils/ppmdi.tar.gz'). 
Lempel-Ziv and grammar-based compressors capture repetitiveness, while the
Burrows-Wheeler one captures only some due to the runs, and the statistical
one is blind to repetitiveness. Then we give the space required by the RLCSA 
alone (which
can count how many times a pattern appears in $T$ but cannot locate the
occurrences nor extract text at random), and RLCSA using a sampling of 512 (the 
minimum space that gives reasonable times for locating and extraction). 
Finally we show the most and least space consuming of our variants over both 
parsings. 

Our least-space variants take 2.5--4.0 times the space of {\tt p7zip}, the 
best LZ77 compressor we know of and the best-performing in our dataset. 
They are also always smaller than 
RLCSA$_{512}$ (up to 6.6 times less) and even competitive with the crippled 
self-index RLCSA-with-no-sampling. The case of {\tt Einstein} is particularly 
illustrative. As it is extremely compressible, it makes obvious how the RLCSA
achieves much compression in terms of the runs of $\Psi$, yet it is unable
to compress the sampling despite many theoretical efforts \cite{MNSV09}. Thus
even a sparse sampling has a very large relative weight when the text is so
repetitive. The data our index needs for locating and extracting, instead, is
proportional to the compressed text size.

\begin{table}[tb]
\begin{center}
\begin{tabular}{l@{~}|@{~}r@{~}||@{~}r@{~}|@{~}r@{~}@{~}|@{~}r@{~}|@{~}r@{~}
	||@{~}r@{~}|@{~}r|@{~}r|@{~}r|@{~}r|@{~}r}
Collection & Size~~ & {\tt p7zip} & {\tt repair} & {\tt bzip2} & {\tt ppmdi} 
	& RLCSA & RLCSA$_{512}$ & LZ77$_5$ & LZ77$_1$ & LZ-End$_5$ & LZ-End$_1$ \\
\hline
{\tt Cere}     & 440MB & 1.14\% & 1.86\% &  2.50\% & 24.09\% 
	& 7.60\% & 8.57\% & 3.74\% & 5.94\% & 6.16\% & 8.96\% \\
{\tt Einstein} & 446MB & 0.07\% & 0.10\% &  5.38\% & 1.61\% 
	& 0.23\% & 1.20\% & 0.18\% & 0.30\% & 0.32\% & 0.48\% \\
{\tt Kernel}   & 247MB & 0.81\% & 1.13\% & 21.86\% & 18.62\% 
	& 3.78\% & 4.71\% & 3.31\% & 5.26\% & 5.12\% & 7.50\% \\
{\tt Leaders}  &  45MB & 1.29\% & 1.78\% &  7.11\% & 3.56\% 
	& 3.32\% & 4.20\% & 3.85\% & 6.27\% & 6.44\% & 9.63\% \\
\end{tabular}
\ \\
\ \\
\caption{Space statistics of our texts.}
\label{tab:space}
\end{center}
\end{table}

\begin{figure}[p]
\centerline{
\includegraphics[angle=-90,width=0.49\textwidth]{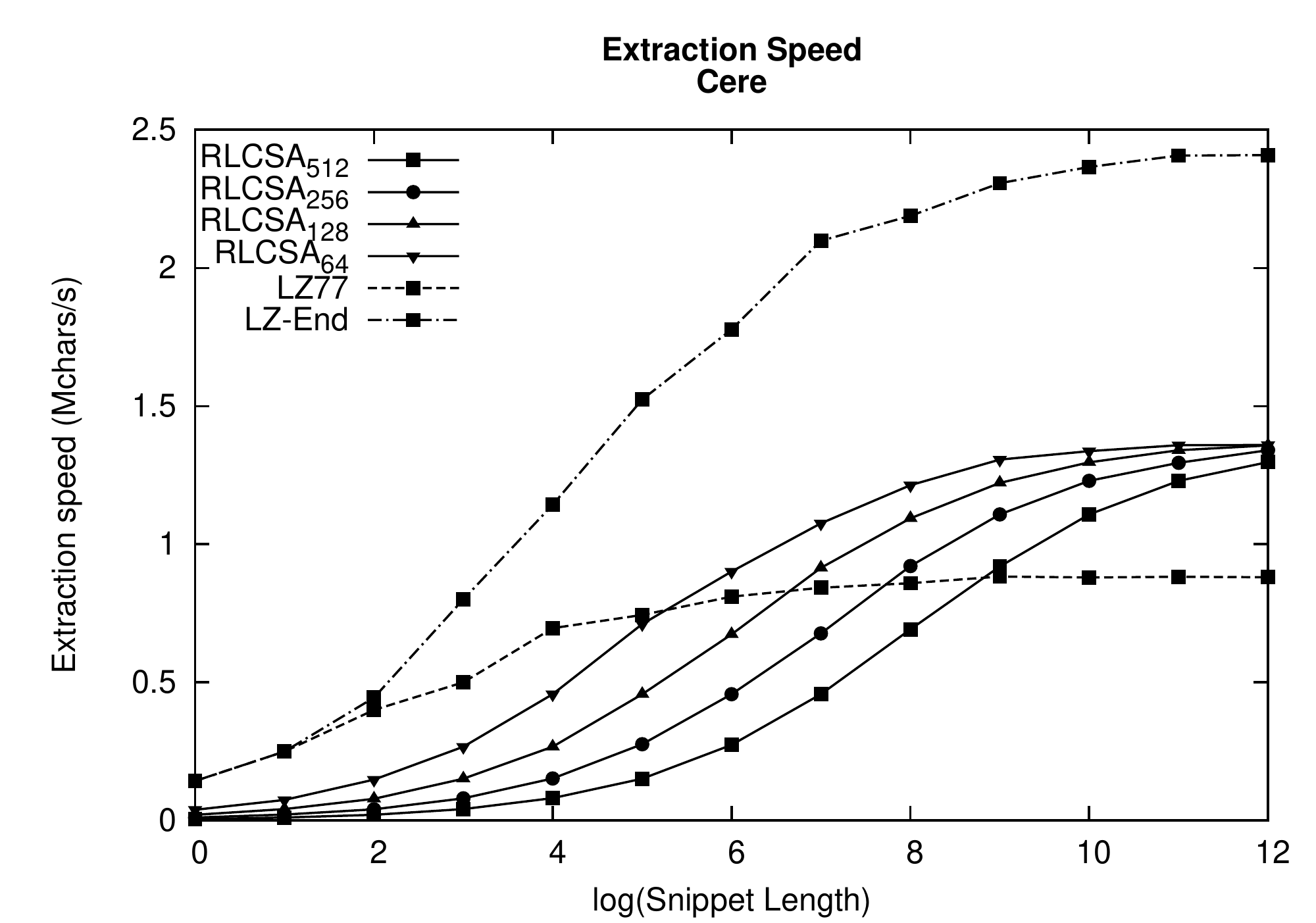}
\includegraphics[angle=-90,width=0.49\textwidth]{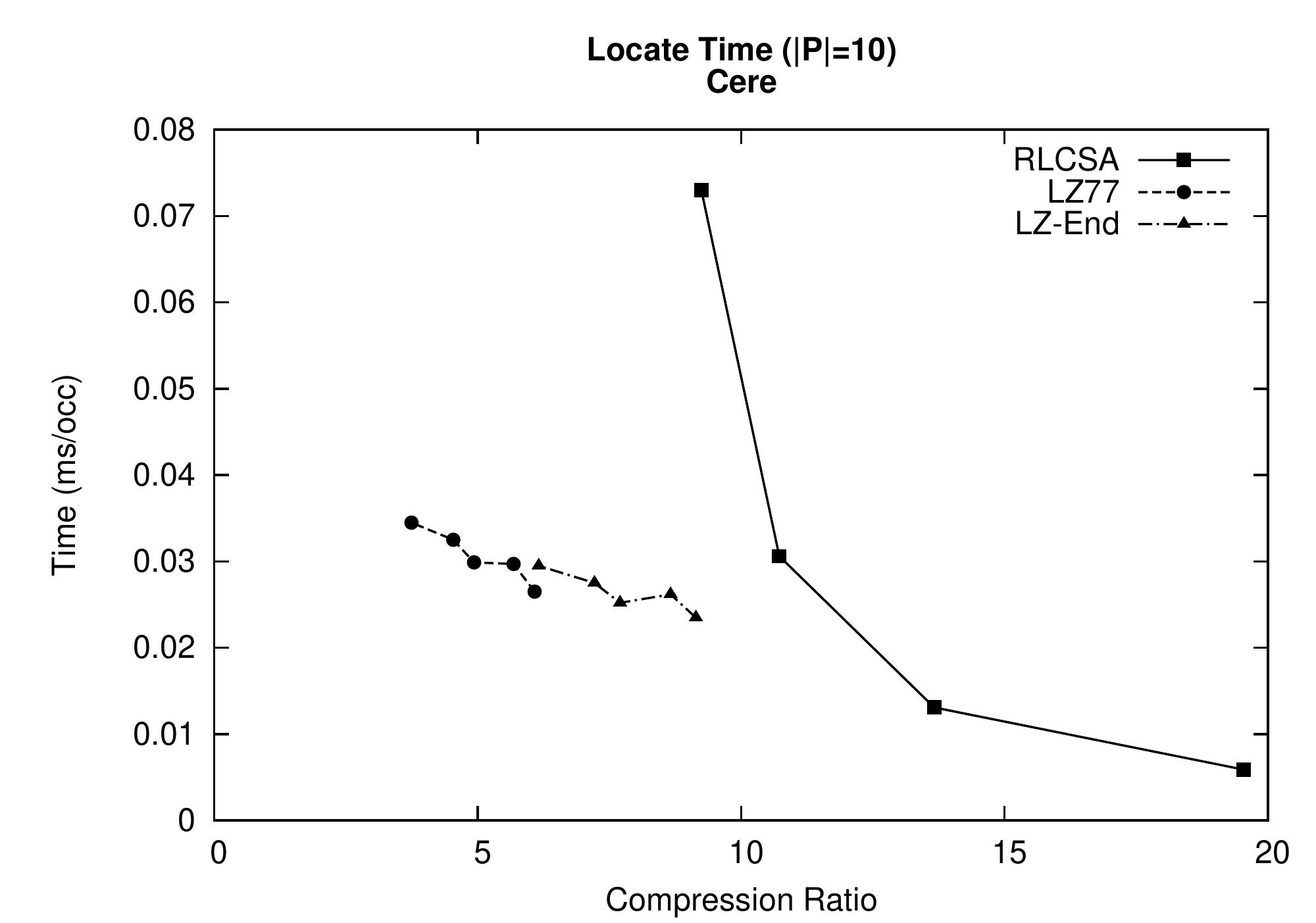}}
\centerline{
\includegraphics[angle=-90,width=0.49\textwidth]{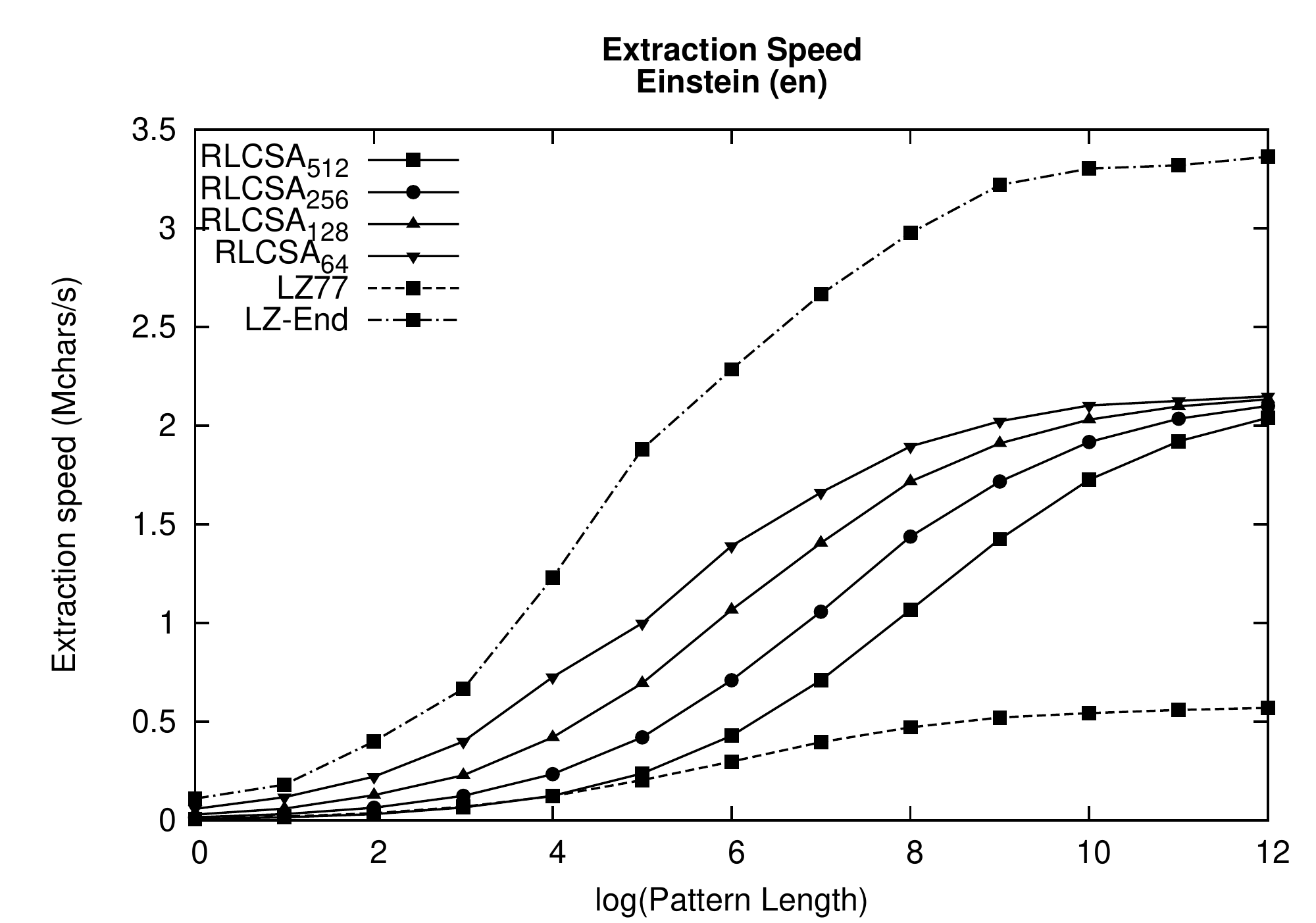}
\includegraphics[angle=-90,width=0.49\textwidth]{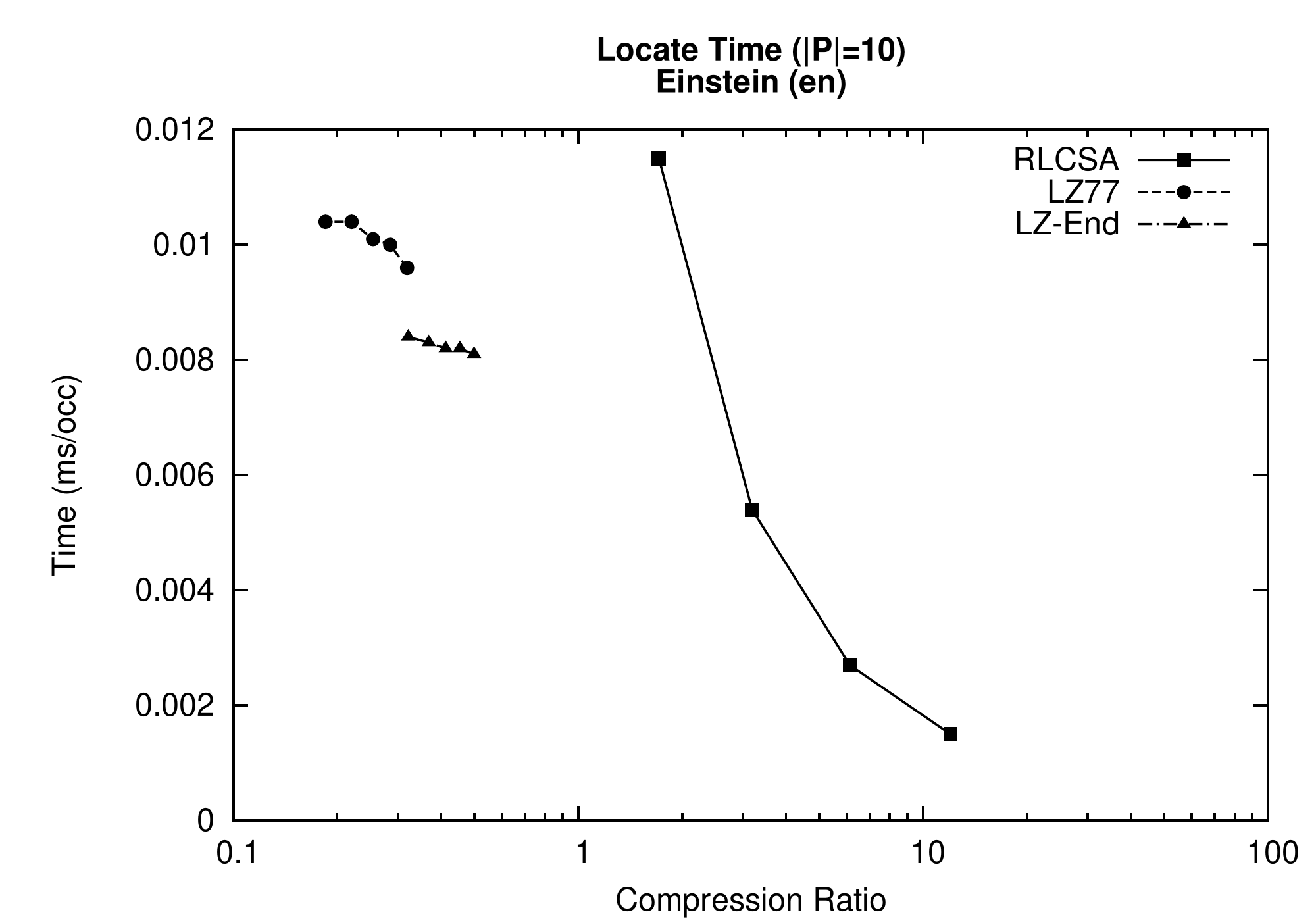}}
\centerline{
\includegraphics[angle=-90,width=0.49\textwidth]{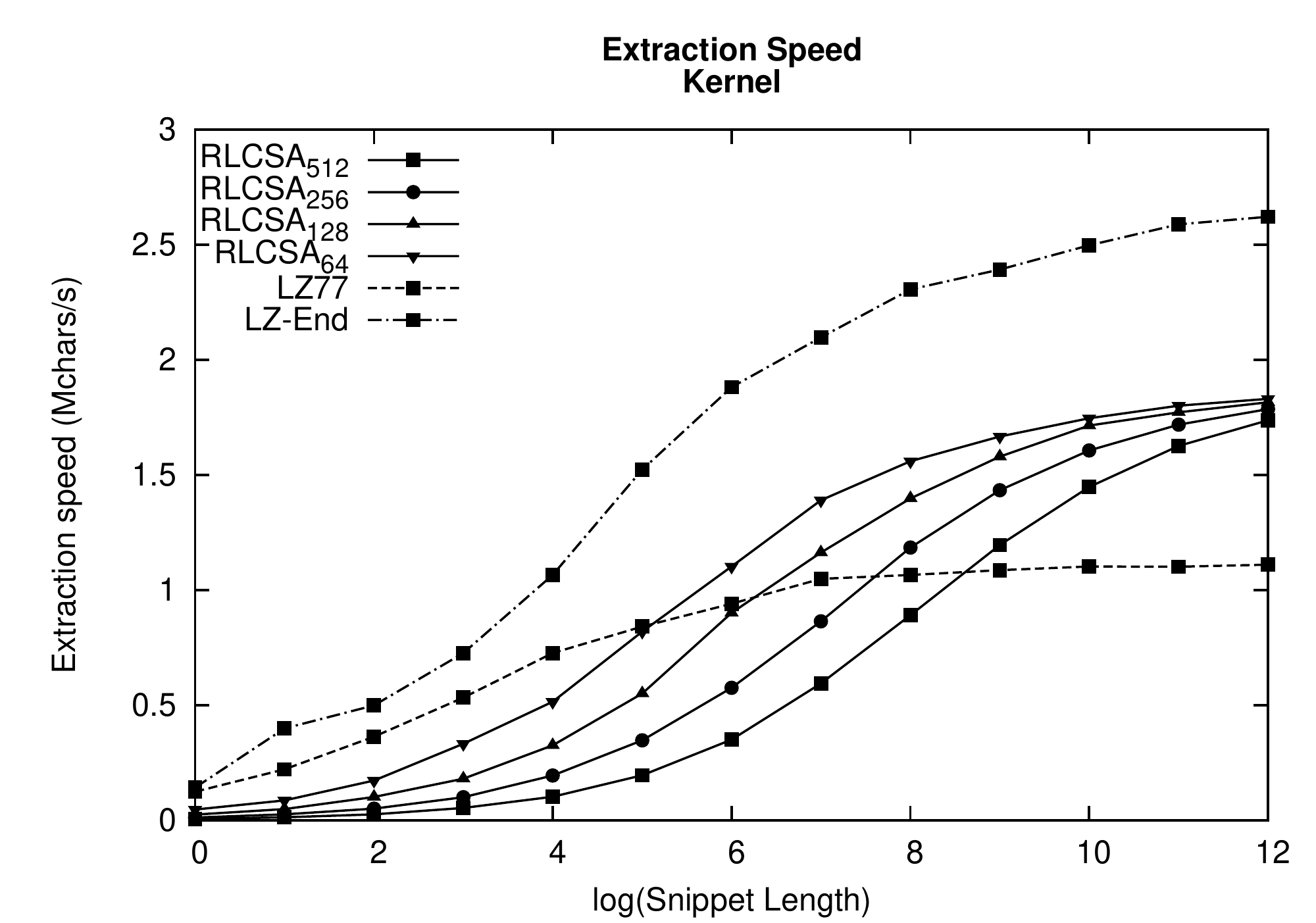}
\includegraphics[angle=-90,width=0.49\textwidth]{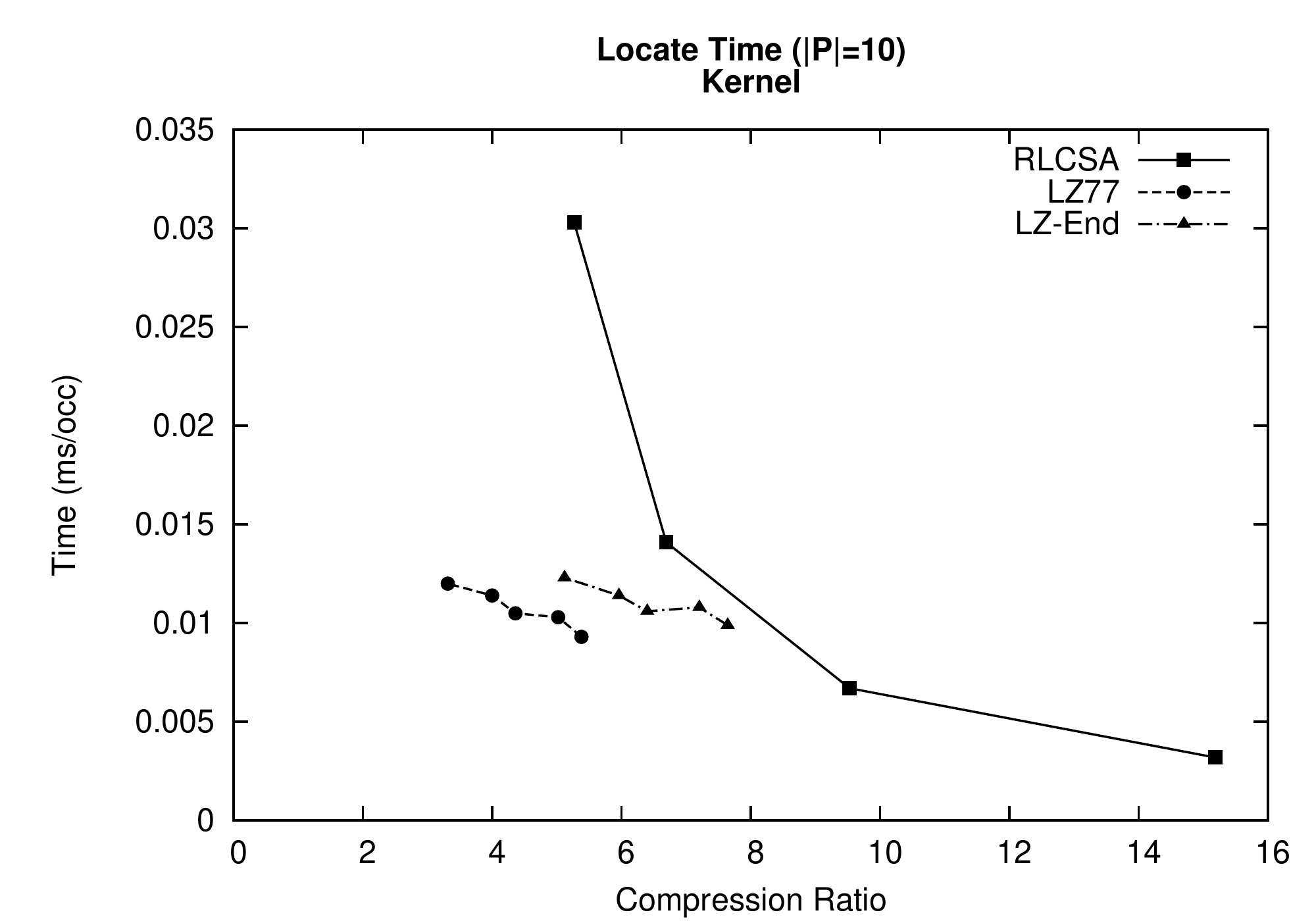}}
\centerline{
\includegraphics[angle=-90,width=0.49\textwidth]{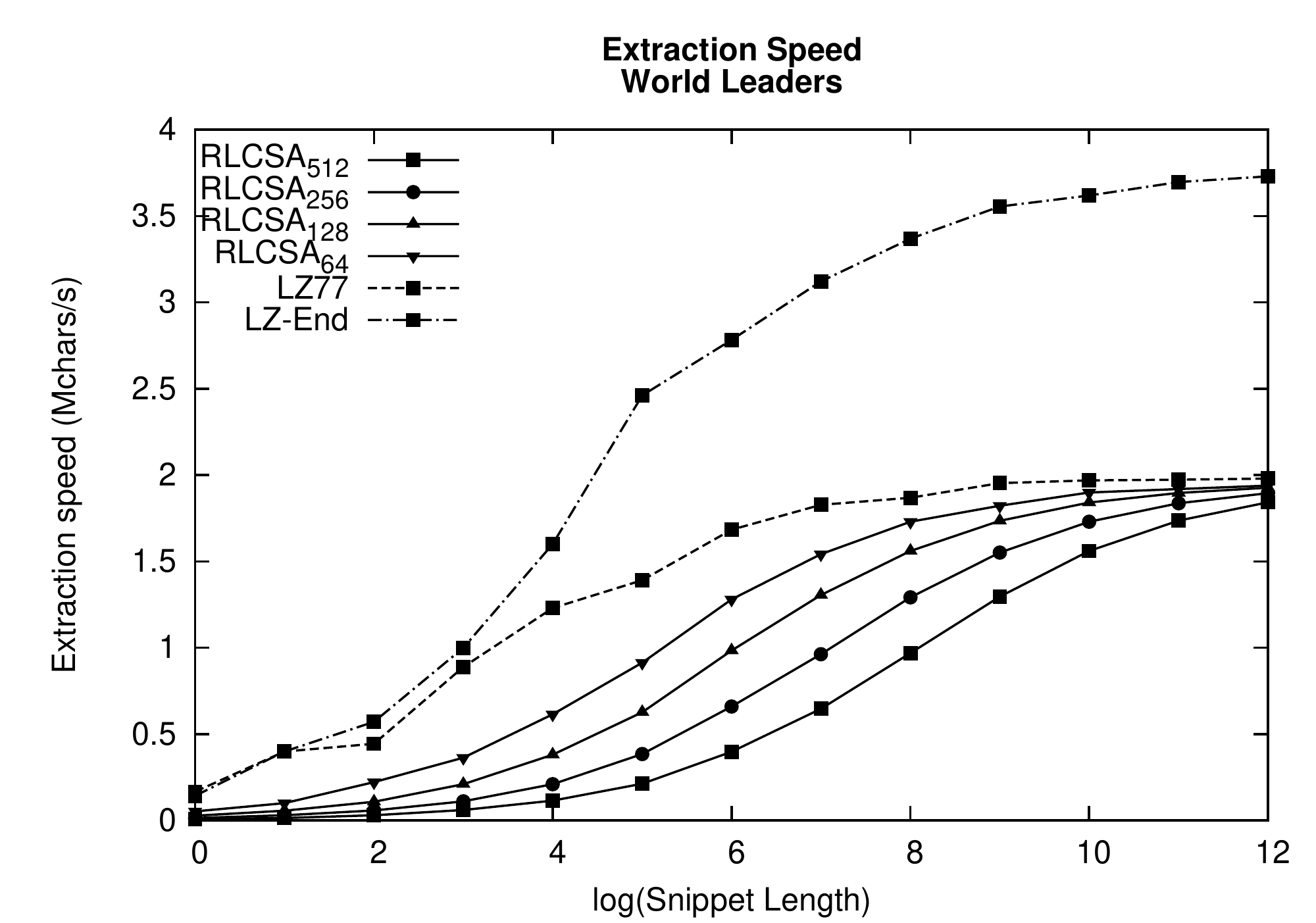}
\includegraphics[angle=-90,width=0.49\textwidth]{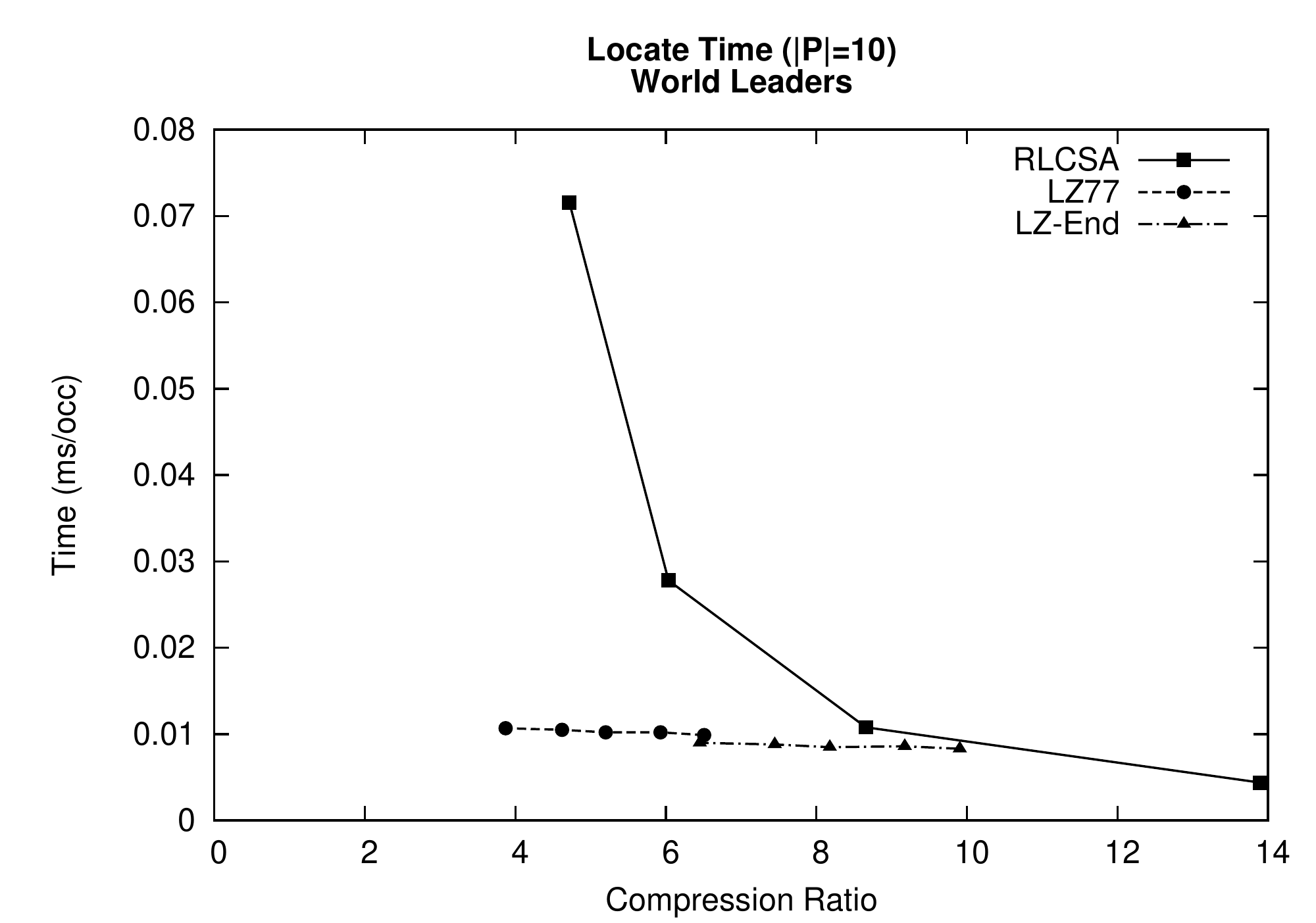}}
\caption{Time performance on the four collections. On the left, extraction
speed as a function of the extracted snippet size (higher is better). On the 
right, time per located occurrence for $m=10$ as a function of the space used
by the index, in percentage of text size (lower and leftwards is better).
On the right the points for RLCSA refer to different sampling rates; for LZ77
and LZ-End refer to the 5 variants (LZ$_5$ is leftmost, LZ$_1$ is rightmost).}
\label{fig:time}
\end{figure}

Fig.~\ref{fig:time} shows times for extracting snippets and for locating
random patterns of length 10. We test RLCSA with various
sampling rates (smaller rate requires more space). It can be seen that our
LZ-End-based index extracts text faster than the RLCSA, while for LZ77 the
results are mixed. For locating, our indexes operate within much less space
than the RLCSA, and are simultaneously faster in several cases. 
See the extended version \cite{Kre10} for more results.

%
%
\section{Conclusions}

We have presented the first self-index based on LZ77 compression, showing
it is particularly effective on highly repetitive text collections, 
which arise in several applications. The new indexes improve upon
the state of the art in most aspects and solve an interesting standing 
challenge. Our solutions to some subproblems, such as that of $prevLess$, 
may be of independent interest.

Our construction needs 6--8 times the original text size and indexes 
0.2--2.0 MB/sec. While this is usual in self-indexes and better than the RLCSA,
it would be desirable to build it within compressed space.

Another important challenge is to be able to restrict the search to a range of 
document numbers, that is, within a particular version, time frame, or version 
subtree. Finally, dynamizing the index, so that at least new text can be added,
would be desirable.

%
%
\bibliographystyle{abbrv}
\bibliography{biblio}
%
\end{document}